\documentclass[a4paper]{article}
\usepackage{ISCSLP2026}
\usepackage{ifthen}
\usepackage{booktabs}   
\usepackage{multirow}
\usepackage{adjustbox}
\usepackage{hyperref}
\usepackage{svg}
\usepackage{xcolor}
\usepackage{graphicx}
\newboolean{blind}
\setboolean{blind}{true}

\title{Preference Optimization with LALM Feedback for Continuous Autoregressive Non-Verbal Vocalization Generation}

\name{
    {
        Jingbin Hu$^1$, Qirui Zhan$^1$, Yuang Cao$^1$, Ziyu Zhang$^1$,  \\ Yunxiang Chen$^2$, Houdun Liu$^2$, Su Feng$^2$, Bengu Wu$^3$, Lei Xie$^{1,\star}$, Liumeng Xue$^{4,5,\star}$\thanks{$^{*}$Corresponding author.}
}
}

\address{
  $^1$Audio, Speech and Language Processing Group (ASLP@NPU),
School of Computer Science, Northwestern Polytechnical University, Xi’an, China \\
$^2$ Shenzhen Pimei Technology Co., Ltd., Guangdong, China \\
$^3$ Yutu Zhineng, Beijing, China \\
$^4$ State Key Laboratory of Novel Software Technology, Nanjing University, Nanjing, China \\
$^5$ School of Intelligence Science and Technology, Nanjing University, Suzhou, China
}

\email{
    {jingbin.hu@mail.nwpu.edu.cn, lxie@nwpu.edu.cn, lmxue@nju.edu.cn, }
}

\begin{document}

\maketitle

\begin{abstract}
We propose a preference optimization framework with Large Audio-Language Model (LALM) feedback for controllable non-verbal vocalization (NVV) generation in continuous autoregressive speech models. To construct preference data without human preference annotation, we build a bilingual prompt corpus by combining NVV-injected real transcripts with LLM-generated semantically aligned prompts, perform stochastic model rollouts, and use a LALM to rank candidate utterances and form same-prompt chosen--rejected pairs. We then adopt a two-stage optimization strategy: Rejection Sampling Fine-Tuning (RSFT) first adapts the model to LALM-selected high-scoring samples, followed by Anchored Flow-DPO, which formulates pairwise preference optimization using utterance-level flow-matching loss and retains the chosen-sample flow-matching objective as an SFT anchor. This design enables DPO-style preference learning without explicit sequence likelihoods while preserving direct supervision on preferred realizations. On the official 1,600-utterance NVVSpeech Challenge Track~2 test set, our method achieves a Final Track2Score of \textbf{75.80} (79.39 ZH / 72.21 EN), outperforming the VoxCPM2 baseline by \textbf{+1.84}. The improvements are mainly driven by higher NVV Accuracy and NVV Perceptual Effect, while Overall Quality remains stable. Speech samples are available~\footnote{Demo:\href{https://hujingbin1.github.io/NVV-TTS-Demo-Page/}{https://hujingbin1.github.io/NVV-TTS-Demo-Page/}}.

\end{abstract}

\noindent\textbf{Index Terms}: speech generation, non-verbal vocalization, preference optimization, rejection sampling fine-tuning, direct preference optimization

\vspace{-15pt}
\section{Introduction}
\vspace{-5pt}

Non-verbal vocalizations (NVVs), such as laughter, sighs, breathing, and crying, convey affective, intentional, and interactional information beyond lexical content and are important for expressive human--computer communication~\cite{xue2026nvvsuperbenchwordsqualitybenchmarkingnonverbal,borisov2025nonverbaltts,liao2025nvspeech,ye2025nonverbalspeech,chen2026mint}. Recent controllable text-to-speech (TTS) systems have made substantial progress in controlling speaking style, emotion, and other expressive attributes~\cite{guo2023prompttts,leng2024prompttts2,zhou2026indextts2,hu2026voicesculptor,chen2026flexivoice,zhu2026omnivoice,qwen2026qwen3tts,zhou2026voxcpm2}. Nevertheless, controllable NVV generation remains challenging because the intended vocalization type and occurrence must be accurately realized while preserving perceptual salience, natural integration with surrounding speech, and overall acoustic quality~\cite{wu2024laughcry,ye2025nonverbalspeech}. The scarcity of diverse, contextually annotated NVV data further limits conventional supervised fine-tuning.

Preference optimization provides a promising alternative for improving generation beyond supervised learning. Direct Preference Optimization (DPO)~\cite{dpo,li2026preference} learns directly from pairwise preferences without requiring a separately trained reward model, and subsequent studies have extended preference optimization from autoregressive language models to diffusion and flow-based generative models~\cite{wallace2024diffusiondpo,liu2025videofeedback,li2026lineardpo}. Preference-based post-training has also been explored for improving the robustness of speech generation~\cite{hu2024rio,yao2025fpo}. However, applying these techniques to continuous autoregressive TTS introduces a fundamental mismatch. Models such as DiTAR, VoxCPM, dots.tts, and VoxCPM2~\cite{jia2025ditar,zhou2025voxcpm,lian2026dotstts,zhou2026voxcpm2} autoregressively generate continuous latent patches through diffusion or flow-matching acoustic heads, rather than discrete tokens with explicit sequence log-likelihoods as assumed by conventional DPO. Moreover, constructing reliable preference pairs for NVVs is non-trivial because candidate utterances must be assessed jointly in terms of NVV correctness, perceptual effectiveness, naturalness, quality, and expression.

To address these challenges, we propose a preference optimization framework with LALM feedback for continuous autoregressive NVV generation. We first construct a bilingual prompt corpus from two complementary sources: random NVV-tag injection into real transcripts yields conflict-rich control conditions, while LLM-generated prompts provide semantically aligned NVV contexts. For each prompt, the baseline model produces multiple stochastic rollouts, which are evaluated by a Large Audio-Language Model (LALM) to automatically construct same-prompt chosen--rejected preference pairs without human annotation. We then adopt a progressive two-stage optimization strategy consisting of Rejection Sampling Fine-Tuning (RSFT), following the reward-ranked fine-tuning paradigm~\cite{dong2023raft}, and Anchored Flow-DPO. RSFT first adapts the model to LALM-selected high-scoring samples. We subsequently adapt Flow-DPO~\cite{liu2025videofeedback} to continuous autoregressive speech generation by using utterance-level flow-matching loss as a surrogate preference signal. Finally, the chosen-sample flow-matching objective is retained as an SFT anchor during Flow-DPO, providing direct supervision on preferred realizations alongside pairwise preference learning.

We evaluate the proposed framework on the official NVVSpeech Challenge\footnote{\href{https://nvvspeech-challenge.github.io/}{https://nvvspeech-challenge.github.io/}} Track~2 test set containing 1,600 utterances. Our proposed system achieves a FinalTrack2Score of \textbf{75.80} (79.39 ZH / 72.21 EN), an absolute improvement of \textbf{+1.84} over the VoxCPM2 baseline (73.96). The improvements are mainly observed in NVV Accuracy and NVV Perceptual Effect while overall acoustic quality remains stable.

\begin{figure*}[th]
    \centering
    \includegraphics[width=0.85\linewidth]{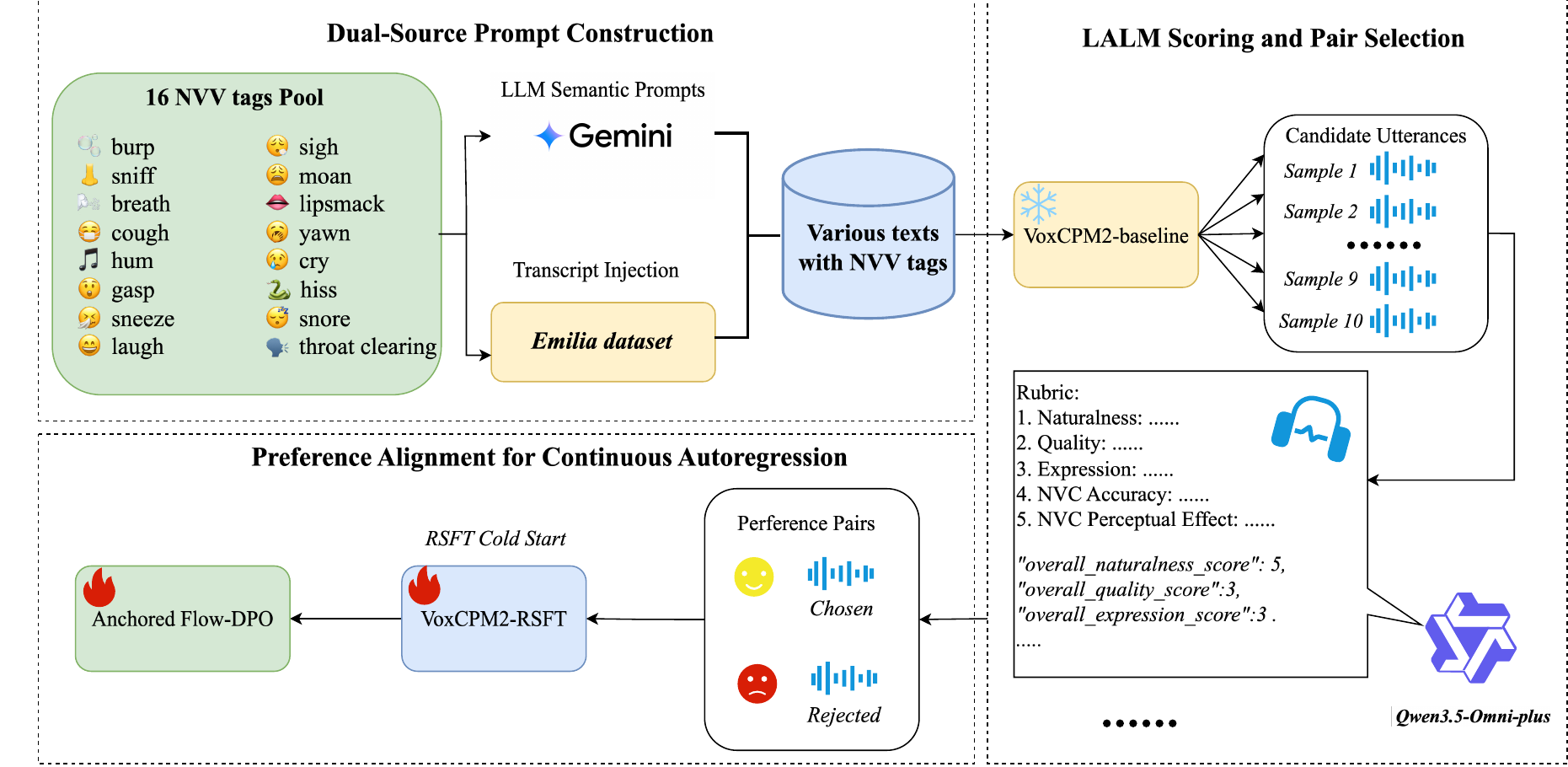}
    \caption{Overview of the proposed LALM-based preference data construction and two-stage preference optimization framework with RSFT and Anchored Flow-DPO.}
    \label{fig:training}
\end{figure*}

Our main contributions are summarized as follows:
\begin{itemize}
    \item We develop an automated bilingual NVV preference-data construction pipeline that combines dual-source prompt construction, stochastic rollouts, and LALM-based multi-dimensional ranking to obtain same-prompt preference pairs without human preference annotation.

    \item We adapt preference optimization to continuous autoregressive speech generation through an utterance-level flow-matching surrogate, and introduce a two-stage optimization strategy combining an RSFT cold start with SFT-anchored Flow-DPO.

    \item Experiments on the official NVVSpeech Challenge Track~2 test set achieve a FinalTrack2Score of \textbf{75.80}, improving the VoxCPM2 baseline by \textbf{+1.84}, with consistent gains in both Chinese and English Track2Scores.
\end{itemize}

\vspace{-15pt}
\section{Method}
\vspace{-5pt}

\label{sec:method}

We propose a preference optimization framework for controllable NVV generation with continuous autoregressive speech models, using LALM feedback to construct preference supervision. As illustrated in Fig.~\ref{fig:training}, we first construct preference pairs through dual-source prompt construction, stochastic rollouts, and LALM-based evaluation, and then progressively optimize the generator using Rejection Sampling Fine-Tuning (RSFT) followed by SFT-anchored Flow-DPO.

\vspace{-5pt}
\subsection{Overview and Continuous Autoregressive Formulation}
\vspace{-5pt}

\label{sec:continuous_ar}

Our base model, VoxCPM2, autoregressively generates continuous VAE latent patches using a language-model backbone and a Local-DiT flow-matching head. Let $c$ denote the text prompt and $X=\{x_m\}_{m=1}^{M}$ an utterance represented by $M$ latent patches. At autoregressive step $m$, the language-model backbone summarizes the text and previous latent patches into a conditional representation $\mu_m=f_{\mathrm{AR}}(c,x_{<m})$. Conditioned on $\mu_m$, the Local-DiT head~\cite{chen2025f5tts,eskimez2024e2tts,wang2025felle} is trained with
\begin{equation}
\mathcal{L}_{\mathrm{FM}}(x_m|\mu_m)
=
\mathbb{E}_{t,\epsilon}
\left[
\left\|
v_\theta(x_{m,t},t,\mu_m)-(\epsilon-x_m)
\right\|_2^2
\right],
\label{eq:patch_fm}
\end{equation}
where $x_{m,t}=(1-t)x_m+t\epsilon$ and $\epsilon\sim\mathcal{N}(0,I)$. At inference, the learned ordinary differential equation (ODE) is integrated backward in flow time from Gaussian noise at $t=1$ toward the data latent at $t=0$.

Although flow matching is trained at the latent-patch level, preference labels are defined over complete generated utterances. We therefore aggregate the patch-level losses into an utterance-level objective:
\begin{equation}
\mathcal{L}_{\mathrm{FM},\theta}(X|c)
=
\frac{1}{M}
\sum_{m=1}^{M}
\mathcal{L}_{\mathrm{FM},\theta}(x_m|\mu_m).
\label{eq:utt_fm}
\end{equation}
Since the flow-matching decoder does not provide the explicit sequence log-likelihoods required by conventional DPO, we use this utterance-level flow-matching loss as a surrogate preference signal.

\vspace{-5pt}
\subsection{LALM-Based Preference Data Construction}
\vspace{-5pt}

\label{sec:preference_data}

\textbf{Dual-Source Prompt Construction.}
Standard SFT data mainly contain semantically consistent NVV examples, providing limited exposure to cases where textual semantics conflict with explicit NVV control tags. We therefore construct 5,000 bilingual Chinese--English prompts from two complementary sources. First, we sample 2,500 authentic transcripts from the Emilia dataset~\cite{he2024emilia} and randomly insert single or multiple NVV tags irrespective of textual semantics, yielding conflict-rich prompts that expose controllability failures under competing textual and tag cues. Second, Gemini 3.1\footnote{\href{https://deepmind.google/models/gemini/pro/}{https://deepmind.google/models/gemini/pro/}} generates another 2,500 prompts in which the textual context is semantically compatible with the target NVV tags.

The two sources play complementary roles in preference mining. Conflict-rich prompts weaken the correlation between textual semantics and explicit NVV tags, making controllability failures easier to expose when the two cues disagree. In contrast, semantically aligned prompts preserve natural context--NVV relationships. Their combination therefore covers both challenging control cases and realistic expressive scenarios.

\textbf{Stochastic Rollouts.}
For each prompt $c$, the baseline model generates $K=10$ candidate utterances $\{X^{(k)}\}_{k=1}^{K}$ using independently sampled Gaussian initializations. Since all candidates share the same textual condition, they capture within-prompt generation variability and enable direct comparison of alternative acoustic realizations under identical NVV control requirements.

\textbf{LALM Scoring and Pair Selection.}
We employ Qwen3.5-Omni-Plus~\cite{qwen3.5-omni} as an automated preference judge. Each candidate is evaluated on five dimensions that mirror the NVVSpeech Challenge Track~2 rubric: NVV Accuracy ($A$), NVV Perceptual Effect ($P$), Overall Naturalness ($N$), Overall Quality ($Q$), and Overall Expression ($E$), each rated on a 1--5 scale. The scores are aggregated as
\begin{equation}
s^{(k)}
=
0.30A^{(k)}
+0.25P^{(k)}
+0.15N^{(k)}
+0.15Q^{(k)}
+0.15E^{(k)}.
\label{eq:judge_score}
\end{equation}
The aggregate score is used only for within-prompt ranking. For each prompt, the highest-scoring and lowest-scoring candidates are selected as the chosen sample $X_w$ and rejected sample $X_l$, respectively, forming the preference dataset
$\mathcal{D}_{\mathrm{pref}}=\{(c,X_w,X_l)\}$.

We use same-prompt extremum pairing to obtain clear preference contrasts while controlling for differences in linguistic content, NVV type, and tag placement. Selecting the highest-scoring and lowest-scoring candidates also avoids potentially ambiguous comparisons between similarly scored middle-ranked samples. Overall, the 5,000 prompts produce 50,000 candidate utterances through $K=10$ rollouts per prompt, yielding 5,000 chosen--rejected preference pairs. The 5,000 chosen utterances are additionally reused as the supervision set for RSFT.

\vspace{-5pt}
\subsection{Preference Alignment for Continuous Autoregression}
\vspace{-5pt}

\label{sec:preference_alignment}

Conventional DPO is not directly applicable to continuous autoregressive speech generation because the flow-matching decoder does not provide the explicit sequence log-likelihoods used in standard DPO. To bridge this mismatch, we adopt a progressive two-stage strategy. Stage~1 performs RSFT on LALM-selected high-scoring samples to obtain a preference-aware initialization. Stage~2 then performs Anchored Flow-DPO, which exploits chosen--rejected comparisons through an utterance-level flow-matching surrogate while retaining direct supervision on the chosen samples.

\textbf{Stage 1: RSFT Cold Start.}
Following the reward-ranked fine-tuning paradigm~\cite{dong2023raft}, we first fine-tune the baseline model on the LALM-selected chosen samples using the standard flow-matching objective:
\begin{equation}
\mathcal{L}_{\mathrm{RSFT}}
=
\mathcal{L}_{\mathrm{FM},\theta}(X_w|c).
\label{eq:rsft}
\end{equation}
Unlike conventional SFT on a fixed corpus, the supervision here is obtained from model-generated candidates: for each prompt, $X_w$ is selected as the highest-scoring realization among $K$ stochastic rollouts by the LALM judge. RSFT therefore converts the LALM-based sample selection into direct supervised learning on high-scoring acoustic realizations.

We use RSFT as a cold start rather than the final alignment objective because it exploits only the chosen samples and does not utilize the relative information carried by the rejected samples. The resulting RSFT checkpoint initializes the trainable policy for Stage~2, while a frozen copy serves as the reference model $\theta_{\mathrm{ref}}$. Thus, subsequent preference optimization is performed relative to the same preference-aware initialization.

\textbf{Stage 2: Anchored Flow-DPO.}
Building on the RSFT initialization, we further exploit the chosen--rejected preference pairs. This stage combines two complementary objectives: a pairwise Flow-DPO term for learning relative preferences and a chosen-sample SFT term for retaining direct supervision on preferred realizations.

\emph{Flow-Matching Preference Objective.}
In standard DPO, preference optimization is expressed through likelihood ratios between the trainable policy and a frozen reference model. Since explicit sequence likelihoods are unavailable for our flow-matching decoder, we construct an analogous relative quantity using the utterance-level flow-matching loss defined in Eq.~\ref{eq:utt_fm}:
\begin{equation}
\Delta_\theta(X,c)
=
\mathcal{L}_{\mathrm{FM},\theta}(X|c)
-
\mathcal{L}_{\mathrm{FM},\mathrm{ref}}(X|c).
\label{eq:fm_gap}
\end{equation}
Here, $\Delta_\theta(X,c)$ measures how the flow-matching loss of utterance $X$ changes under the trainable policy relative to the frozen reference model. A smaller $\Delta_\theta(X,c)$ means that the utterance receives a lower relative flow-matching loss under the trainable policy.

Given a same-prompt preference pair $(X_w,X_l)$, where $X_w$ and $X_l$ denote the chosen and rejected utterances, respectively, we optimize
\begin{equation}
\mathcal{L}_{\mathrm{DPO}}
=
-\log\sigma
\left(
\beta
\left[
\Delta_\theta(X_l,c)
-
\Delta_\theta(X_w,c)
\right]
\right),
\label{eq:flow_dpo}
\end{equation}
where $\sigma(\cdot)$ is the sigmoid function and $\beta$ controls the preference sharpness. Minimizing this objective encourages
$\Delta_\theta(X_w,c)<\Delta_\theta(X_l,c)$, so that the trainable policy favors the chosen utterance over the rejected one relative to the same reference model. Because both utterances correspond to the same prompt, the comparison focuses on alternative acoustic realizations under identical linguistic and NVV control conditions.

\emph{SFT Anchor.}
The pairwise Flow-DPO objective specifies only a relative ordering between $X_w$ and $X_l$. In particular, the preference margin can be enlarged either by decreasing the relative loss of the chosen sample or by increasing that of the rejected sample; it does not explicitly require the model to maintain a low flow-matching loss on $X_w$. We therefore retain the chosen-sample flow-matching objective during preference optimization:
\begin{equation}
\mathcal{L}
=
\mathcal{L}_{\mathrm{DPO}}
+
\gamma
\mathcal{L}_{\mathrm{FM},\theta}(X_w|c),
\label{eq:anchored_dpo}
\end{equation}
where $\gamma$ controls the strength of the SFT anchor. The two terms provide complementary supervision: $\mathcal{L}_{\mathrm{DPO}}$ learns relative preferences from chosen--rejected pairs, whereas the SFT term explicitly encourages the policy to fit the LALM-selected high-scoring samples. RSFT introduces this chosen-sample supervision before preference optimization, while the SFT anchor retains it throughout Flow-DPO, yielding the proposed two-stage Anchored Flow-DPO framework.

\begin{table*}[th]
\centering
\caption{Main results and ablations on the official NVVSpeech Challenge Track~2 test set (1,600 utterances). 
A/P/N/Q/E denote the raw 1--5 ratings for NVV Accuracy, NVV Perceptual Effect, Naturalness, Quality, and Expression, respectively. 
Track2 denotes the official language-specific Track2Score, and Final averages the ZH and EN scores. 
Best results in each column are shown in bold.}
\label{tab:main_ablation}

\footnotesize
\setlength{\tabcolsep}{3.6pt}
\renewcommand{\arraystretch}{1.15}

\begin{adjustbox}{max width=\textwidth,center}
\begin{tabular}{lcccccc|cccccc|c}
\toprule
\multirow{2}{*}{\textbf{Model / Strategy}}
& \multicolumn{6}{c|}{\textbf{Chinese (ZH)}}
& \multicolumn{6}{c|}{\textbf{English (EN)}}
& \multirow{2}{*}{\textbf{Final} $\uparrow$} \\
\cmidrule(lr){2-7}
\cmidrule(lr){8-13}
& A $\uparrow$
& P $\uparrow$
& N $\uparrow$
& Q $\uparrow$
& E $\uparrow$
& Track2 $\uparrow$
& A $\uparrow$
& P $\uparrow$
& N $\uparrow$
& Q $\uparrow$
& E $\uparrow$
& Track2 $\uparrow$
& \\
\midrule

VoxCPM2 (Baseline)
& 4.52 & 4.41 & 3.73 & 3.02 & 4.22 & 77.62
& 4.37 & 3.90 & 3.37 & 3.08 & 3.74 & 70.30
& 73.96 \\

RSFT (Cold Start)
& 4.57& 4.43& 3.78& 3.04& 4.26& 78.54
& 4.47& 3.90& 3.42& 3.10& 3.77& 71.46
& 75.00 \\

Flow-DPO (w/o SFT anchor)
& 4.58& 4.42& 3.79& \textbf{3.06}& 4.30& 78.73
& \textbf{4.51}& 3.96&3.41&3.10&3.78&72.12 
& 75.43 \\

RSFT + Flow-DPO (w/o SFT anchor)
&4.58&\textbf{4.50}&\textbf{3.84}&3.04&4.30&\textbf{79.41}
&4.46&3.93&3.44&\textbf{3.11}&\textbf{3.83}&71.97
& 75.69 \\

\textbf{RSFT + Flow-DPO (w/ SFT anchor, proposed)}
& \textbf{4.60} & 4.48 & 3.82 & 3.05 & \textbf{4.31} & 79.39
& 4.45 & \textbf{3.99} & \textbf{3.45} & \textbf{3.11}& 3.81 & \textbf{72.21}
& \textbf{75.80} \\

\bottomrule
\end{tabular}
\end{adjustbox}
\end{table*}

\begin{table}[htbp]
\centering
\caption{Subjective evaluation results. Metrics A/P/N/Q/E are consistent with Table 1.}
\label{tab:subjective_evaluation_single}
\footnotesize
\setlength{\tabcolsep}{2.5pt} 
\renewcommand{\arraystretch}{1.1}

\begin{tabular}{lcccc}
\toprule
\multirow{2}{*}{\textbf{Metric}} & \multicolumn{2}{c}{\textbf{Chinese (ZH)}} & \multicolumn{2}{c}{\textbf{English (EN)}} \\
\cmidrule(lr){2-3} \cmidrule(lr){4-5}
& \textbf{Baseline} & \textbf{Proposed} & \textbf{Baseline} & \textbf{Proposed} \\
\midrule
A & $2.45 \pm 0.20$ & $\mathbf{2.60} \pm 0.19$ & $2.89 \pm 0.19$ & $\mathbf{3.00} \pm 0.19$ \\
P & $1.98 \pm 0.17$ & $\mathbf{2.11} \pm 0.16$ & $2.68 \pm 0.17$ & $\mathbf{2.78} \pm 0.17$ \\
N & $3.03 \pm 0.09$ & $\mathbf{3.06} \pm 0.09$ & $3.33 \pm 0.08$ & $\mathbf{3.40} \pm 0.08$ \\
Q & $3.29 \pm 0.09$ & $\mathbf{3.33} \pm 0.08$ & $3.42 \pm 0.10$ & $\mathbf{3.57} \pm 0.10$ \\
E & $2.83 \pm 0.10$ & $\mathbf{2.90} \pm 0.09$ & $3.05 \pm 0.09$ & $\mathbf{3.18} \pm 0.10$ \\
\midrule
Overall & $2.71 \pm 0.09$ & $\mathbf{2.80} \pm 0.09$ & $3.07 \pm 0.09$ & $\mathbf{3.19} \pm 0.09$ \\
\bottomrule
\end{tabular}
\end{table}

\vspace{-10pt}
\section{Experiments}
\vspace{-5pt}

\subsection{Experimental Setup}
\vspace{-5pt}

\textbf{Dataset and Evaluation.} 
We evaluate our framework on the official NVVSpeech Challenge Track~2 test set, comprising 1,600 utterances (800 Chinese and 800 English). Evaluation follows the official Track~2 protocol using Gemini 2.5 Pro~\cite{gemini2025gemini25} over the same five dimensions and weights described in Sec.~\ref{sec:preference_data}. For each language, the official Track2Score is computed from the weighted component scores, and the final bilingual score averages the Chinese and English results. Notably, the evaluation model differs from Qwen3.5-Omni-Plus used for preference construction, avoiding direct optimization against the same LALM used for testing.

\textbf{Implementation Details.} 
We implement our framework on the official VoxCPM2 Track~2 baseline\footnote{\href{https://github.com/NVVSpeech-Challenge/NVVSpeech-Challenge-Track2-Baseline}{NVVSpeech Challenge Track~2 Baseline}}. RSFT is trained for 1 epoch, followed by Flow-DPO for 3 epochs. For the proposed RSFT-initialized variants, the RSFT checkpoint initializes the trainable policy, while a frozen copy serves as the reference model $\theta_{\mathrm{ref}}$. For the direct Flow-DPO ablation without RSFT, the VoxCPM2 baseline is used correspondingly as both the policy initialization and the frozen reference. We use a batch size of 1 with 16 gradient-accumulation steps. Optimization uses AdamW with a learning rate of $1\times10^{-6}$, cosine decay with warmup, and gradient clipping at 1.0. The preference sharpness $\beta$ and SFT-anchor weight $\gamma$ are both set to 1.0.

\vspace{-10pt}
\subsection{Main Results}
\vspace{-5pt}

Table~\ref{tab:main_ablation} reports the results on the official NVVSpeech Challenge Track~2 test set. The VoxCPM2 baseline achieves a FinalTrack2Score of 73.96, while our proposed system reaches \textbf{75.80}, yielding an absolute improvement of \textbf{+1.84}. Improvements are observed in both languages, from 77.62 to \textbf{79.39} for Chinese and from 70.30 to \textbf{72.21} for English.

The gains are primarily concentrated in the NVV-specific dimensions. Compared with the baseline, NVV Accuracy ($A$) improves from 4.52 to \textbf{4.60} and NVV Perceptual Effect ($P$) from 4.41 to \textbf{4.48} for Chinese. For English, $A$ increases from 4.37 to \textbf{4.45} and $P$ from 3.90 to \textbf{3.99}. Meanwhile, Overall Quality ($Q$) remains nearly unchanged (3.02$\rightarrow$3.05 for ZH and 3.08$\rightarrow$3.11 for EN), indicating that the gains in NVV controllability and perceptual effectiveness are achieved without degrading the overall quality score.

As shown in Table~\ref{tab:subjective_evaluation_single}, the human subjective evaluation, conducted by 12 Chinese and 12 English native listeners, further validates the overall effectiveness of our proposed approach. Our proposed model consistently outperforms the baseline across all evaluation metrics ($A$, $P$, $N$, $Q$, and $E$) in both Chinese and English test sets. Notably, the overall preference score achieves substantial improvements, rising from $2.71$ to $2.80$ in Chinese and $3.07$ to $3.19$ in English. These across-the-board gains demonstrate that our preference optimization strategy effectively boosts task-specific accuracy and perception without sacrificing naturalness or speech quality, confirming its comprehensive advantage and cross-lingual robustness.

\vspace{-10pt}
\subsection{Ablation Study}
\vspace{-5pt}

Table~\ref{tab:main_ablation} further examines the contributions of RSFT, Flow-DPO, and the SFT anchor. RSFT alone improves the Final score from 73.96 to 75.00 (+1.04 gain), providing consistent performance boosts across both Chinese (77.62 $\rightarrow$ 78.54) and English (70.30 $\rightarrow$ 71.46). This confirms that winner-only fine-tuning serves as an effective cold-start strategy.

Applying Flow-DPO directly to the baseline without the SFT anchor increases the Final score to 75.43, demonstrating the power of pairwise preference optimization. Notably, direct Flow-DPO achieves a larger improvement over the baseline than RSFT alone (+1.47 vs.\ +1.04). While RSFT exploits only the selected winning utterances, Flow-DPO additionally leverages the relative contrast between chosen and rejected samples, providing richer supervision beyond winner-only fine-tuning.

Initializing Flow-DPO from the RSFT checkpoint further elevates the Final score from 75.43 to 75.69 (+0.26 gain). This combination is particularly beneficial for Chinese, pushing its Track2 score from 78.73 to 79.41 and significantly enhancing the NVV Perceptual Effect ($P$) from 4.42 to 4.50. This indicates that RSFT provides a superior starting distribution for subsequent preference alignment.

Finally, incorporating the SFT anchor achieves the best overall Final score of \textbf{75.80}. The SFT anchor effectively prevents policy drift, notably recovering and boosting English performance (Track2: 71.97 $\rightarrow$ \textbf{72.21}; $P$: 3.93 $\rightarrow$ \textbf{3.99}) while maintaining top-tier Chinese results (achieving the highest NVV Accuracy $A$ of \textbf{4.60}). Overall, the ablations confirm that pairwise Flow-DPO drives the primary preference gains, while RSFT initialization and SFT anchoring play complementary roles in stabilizing optimization and balancing cross-lingual quality.
\vspace{-17.5pt}
\section{Conclusion}
\vspace{-5pt}
We present an LALM-guided preference optimization framework for controllable NVV generation in continuous autoregressive speech models, adopting a two-stage strategy of RSFT cold start followed by SFT-anchored Flow-DPO. By leveraging utterance-level flow-matching loss, it enables likelihood-free DPO-style alignment while retaining direct supervision on preferred samples. On the NVVSpeech Challenge Track~2 test set, our system reaches a FinalTrack2Score of \textbf{75.80} (+\textbf{1.84} over VoxCPM2 baseline), significantly improving NVV accuracy and perceptual quality.

\bibliographystyle{IEEEtran}

\bibliography{mybib}


\end{document}